\documentstyle[psfig,conf_iap,]{article}
\newcommand{\Lya}{Ly$\alpha$}

\newcommand{\cm}{cm$^{-2}$}
\newcommand{\kms}{km s$^{-1}$}

\begin{document}
\heading{%
%
Coincidences of High Density Peaks in UVES Spectra of QSO Pairs  
%
} 
\par\medskip\noindent
\author{%
Valentina D'Odorico$^1$, Patrick Petitjean$^1$, Stefano
Cristiani$^2$ 
}
\address{%
Institut d'Astrophysique de Paris, 98bis Boulevard Arago, 
       F--75014 Paris, France}
\address{
European Southern Observatory, Karl-Schwarzschild-Strasse
2, D--85748 Garching, Germany 
}

\begin{abstract}
We present preliminary results of an investigation of
the clustering properties of high matter density peaks
between redshift $\sim 2$ and $\sim 3$, as
traced by Lyman limit and Damped \Lya\ systems in
spectra of close QSO pairs and groups. 

\end{abstract}
\section{Introduction}
Paired lines of sight (LOS) toward high redshift quasars,
with angular separations up to a few arcminutes, are a
useful tool to investigate the clustering properties of
absorption lines. 
We have obtained with UVES high resolution spectra
($R\simeq 37000$) of two QSO pairs with separations 1 and
5 arcmin and a QSO triplet with reciprocal separations 1,
8 and 8 arcmin, spanning the redshift range $1.6 \lsim z
\lsim 3.2$ (see \cite{vale01} for further details). 
We assume that high matter density peaks
are traced by optically thick absorbers
(i.e. with column density $N({\rm HI}) \gsim 2\times
10^{17}$ \cm).
The present spectra are scanned to detect the presence of 
high density peaks. We find 5 systems with $N({\rm HI})
\gsim 10^{19}$ \cm\ and 7 with $2\times 10^{17}\lsim\, 
N({\rm HI}) \lsim 10^{19}$ \cm.  
As a second step, we look systematically for
coincident absorptions at the same redshift as the
identified high column density systems. 

\section{Results}
%
Out of 5 detected absorption systems with $N({\rm HI})
\gsim 10^{19}$ \cm, 3 of them have a corresponding metal
system in the companion LOS at a velocity difference of
less than 200 \kms. One of them is at less then 1000
\kms\ from the emission redshift of the paired QSO
(also marking a high density peak) and the last
one has a corresponding, weak \Lya\ absorption line but
no metal absorption within $\sim 9000$ \kms. 
From the number density of C~IV systems with rest
equivalent width $w_0 > 0.15$ \AA, as a function of
redshift \cite{steidel90}, we compute the chance
probability (in the hypothesis of null clustering) to
detect a C~IV absorption line within 200 \kms\ between $z
=2$ and 3, ${\cal P}(z) \simeq 0.001$.  
The transverse spatial separation over which these
coincidences happen varies between $\sim 4$ and 7
$h_{100}^{-1}$  comoving Mpc, which suggests that we are
detecting the clustering signal of galactic objects, as
verified in the past 
\cite{fh93,francis96,fwd97}.  
These separations could be indeed reasonable correlation 
lengths for normal or dwarf galaxies at this redshift
since at $z \sim 3$ the Lyman break galaxies are found to
show correlation lengths $\sim 2\ h_{100}^{-1}$  Mpc 
\cite{giav98,cris01}.
 
As for the 7 Lyman limit systems (LLS) with $N({\rm HI})
\lsim 10^{19}$ \cm, 4 form two coincident pairs 
along two LOS separated by $\sim 1$ arcmin; the
similarity of the coincident absorptions suggests that
the two LOS could be piercing a coherent filament-like
structure. 
The remaining three systems do show corresponding \Lya\
absorption lines but no metal absorption within 3000
\kms, at transverse spatial separations of $\sim 3.8\
h_{100}^{-1}$ Mpc, and in the triplet $\sim 6.7\
h_{100}^{-1}$ Mpc and $\sim 840\ h_{100}^{-1}$ kpc.

%
%
%
\begin{figure}
\centerline{\vbox{
\psfig{figure=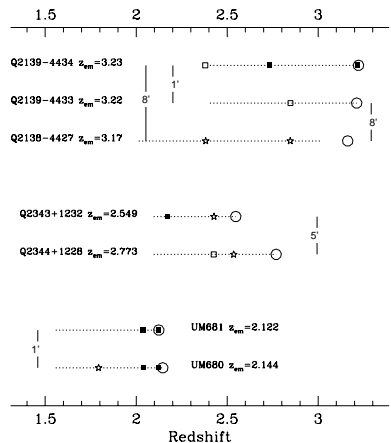,height=6.cm}}}
\caption[]{Summary of the observed coincidences as a
function of redshift. 
The dotted lines mark the observed redshift ranges. The  
angular separations of the quasars are reported between  
the solid vertical lines. The
symbols are: open square for metal systems, solid square
for LLS with $N({\rm HI})<10^{19}$ \cm\ and star for
system with $N({\rm HI})>10^{19}$ \cm. 
The big open circles mark the emission
redshift of the quasars}
\end{figure}

\acknowledgements{V.D. is supported by a Marie Curie
individual fellowship  from the European Commission under 
the programme ``Improving Human Research Potential and
the Socio-Economic Knowledge Base'' (Contract
no. HPMF-CT-1999-00029). We thank
C. Ledoux for the UVES spectrum of Q2138-4427.}

\vfill
\end{document}